\documentclass[preprint]{acmart}

\usepackage{graphicx}
\usepackage{booktabs} 
\usepackage{amsmath}
\usepackage{algorithm,algpseudocode}

\usepackage{fix-cm}
\usepackage[T1]{fontenc}

\usepackage[belowskip=0pt,aboveskip=0pt]{caption}

\usepackage{titlesec}

\AtBeginDocument{%
  \providecommand\BibTeX{{%
    \normalfont B\kern-0.5em{\scshape i\kern-0.25em b}\kern-0.8em\TeX}}}

\copyrightyear{2020}
\acmYear{2020}
\setcopyright{acmcopyright}\acmConference[UMAP '20]{Proceedings of the 28th ACM Conference on User Modeling, Adaptation and Personalization}{July 14--17, 2020}{Genoa, Italy}
\acmBooktitle{Proceedings of the 28th ACM Conference on User Modeling, Adaptation and Personalization (UMAP '20), July 14--17, 2020, Genoa, Italy}
\acmPrice{15.00}
\acmDOI{10.1145/3340631.3394884}
\acmISBN{978-1-4503-6861-2/20/07}

\begin{document}
\fancyhead{} 
\title{Keen2Act: Activity Recommendation in Online Social Collaborative Platforms}


\author{Roy Ka-Wei Lee}
\affiliation{%
  \institution{University of Saskatchewan, Canada}
 }
\email{roylee@cs.usask.ca}

\author{Thong Hoang}
\affiliation{%
  \institution{Singapore Management University, Singapore}
 }
\email{vdthoang.2016@smu.edu.sg}

\author{Richard J. Oentaryo}
\affiliation{%
  \institution{McLaren Applied, Singapore}
 }
\email{richard.oentaryo@mclaren.com}

\author{David Lo}
\affiliation{%
  \institution{Singapore Management University, Singapore}
 }
\email{davidlo@smu.edu.sg}



\begin{abstract}
Social collaborative platforms such as GitHub and Stack Overflow have been increasingly used to improve work productivity via collaborative efforts. To improve user experiences in these platforms, it is desirable to have a recommender system that can suggest not only items (e.g., a GitHub repository) to a user, but also activities to be performed on the suggested items (e.g., forking a repository). To this end, we propose a new approach dubbed \textsf{Keen2Act}, which decomposes the recommendation problem into two stages: the \textit{Keen} and \textit{Act} steps. The \textit{Keen} step identifies, for a given user, a (sub)set of items in which he/she is likely to be interested. The \textit{Act} step then recommends to the user which activities to perform on the identified set of items. This decomposition provides a practical approach to tackling complex activity recommendation tasks while producing higher recommendation quality. We evaluate our proposed approach using two real-world datasets and obtain promising results whereby \textsf{Keen2Act} outperforms several baseline models.
\end{abstract}




\maketitle

\section{Introduction}
\label{sec:introduction}
Users are increasingly adopting social collaborative platforms for collaborative activities. GitHub and Stack Overflow are two such popular platforms; GitHub is a collaborative software development platform that allows code sharing and version control, while Stack Overflow is a technical question-and-answer community-based website. As these social collaborative platforms gain popularity, many research studies have proposed recommender systems to improve the usability of these platforms. For example, there are works that recommend Stack Overflow questions for users to answer \cite{wang2015,wang2016}. Similarly, in GitHub, researchers have proposed methods to recommend relevant repositories to a user \cite{guendouz2015,jiang2017}. 

Many of the existing works, however, focus largely on recommending either items or a single type of activity to users, ignoring the fact that the users can perform multiple types of activity on these platforms. For example, GitHub users may \textit{fork} or \textit{watch} repositories and Stack Overflow users may \textit{answer} or \textit{favorite} questions. 
Recommending multiple types of activities to a user is a challenging task. A na\"ive solution would be to recommend individual activities separately, treating them as independent tasks. However, there might be insufficient observations for learning the user activities at such granularity. Another possible solution is to simply learn the different types of activities together, but such solution creates sparsity issue, having to learn for all possible item-activity pairs.

To tackle these challenges, we propose a new recommendation approach called \textsf{Keen2Act}\footnote{Source code: https://gitlab.com/bottle\_shop/scp/keen2act}, which learns the users' item-level and activity-level interests in a step-wise manner to achieve (joint) item-activity recommendation. In particular, the main contribution of \textsf{Keen2Act} is that it features a novel two-stage process that decomposes the activity recommendation problem into predicting a user's interests at item level and subsequently predicting at activity level. To the best of our knowledge, this is also the first work that achieves multi-typed activity recommendation in social collaborative platforms. Finally, empirical studies on real-world GitHub and Stack Overflow datasets have shown promising results whereby the proposed approach outperforms several baseline methods.

\begin{figure*}[!t]
\includegraphics[width=1.0\textwidth]{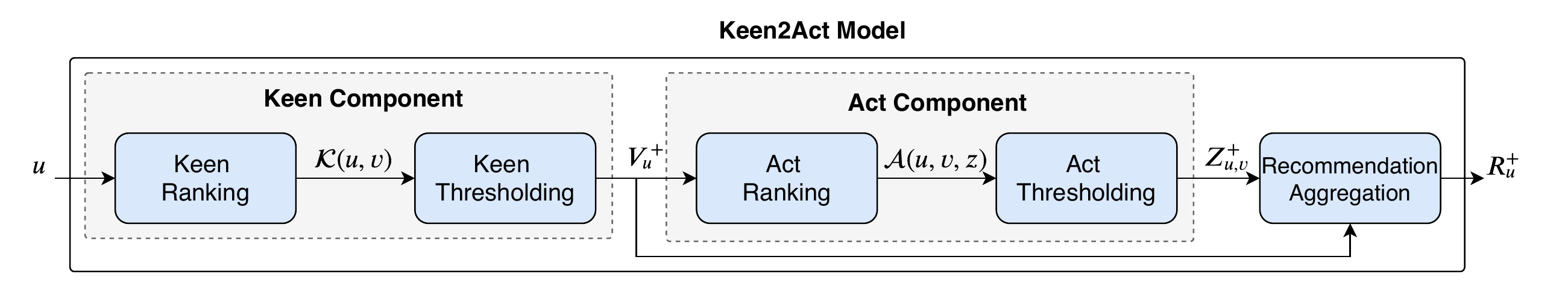}
\caption{\textsf{Keen2Act} Framework}
\label{fig:framework}
\end{figure*}

\section{Related Work}
\label{sec:related}

Research studies on prediction and recommendation in social collaborative platforms broadly fall into two categories: (i) finding experts to perform certain platform tasks \cite{Riahi2012,Yang2013,choetkiertikul2015,xu2016,wang2017,huang2017,Allaho2014,yu2016,rahman2016} and (ii) recommending items to users in a platform \cite{deSouza2014,wang2016,guendouz2015,zhang2014,jiang2017,wang2015}. Under category (i), the works on Stack Overflow mainly involve devising methods to find experts to answer questions \cite{Riahi2012,Yang2013,choetkiertikul2015,xu2016,wang2017}, while for GitHub a user is identified as an expert if (s)he reviews pull-requests and code for repositories \cite{yu2016,rahman2016}. The works under category (ii) largely focus on recommending items to users, without specifying activities to be performed on the items. For example, several works aim to recommend relevant Stack Overflow posts \cite{deSouza2014,wang2015} and Github repositories \cite{zhang2014,jiang2017} to users. 

Our work deviates from the existing works under category (ii) in several ways. Firstly, to the best of our knowledge, \textsf{Keen2Act} constitutes the first work that focuses on recommending not only items but also specific activities under each recommended item. Our approach has also been applied to more than one platform (i.e., GitHub and Stack Overflow).

\section{Proposed Approach}
\label{sec:model}

\subsection{Problem Formulation}

We define the activity recommendation problem as follows: \emph{For a given user, which items should (s)he choose and what activities should (s)he perform on those items}? The proposed \textsf{Keen2Act} approach addresses this joint item-activity recommendation problem by breaking it down into two sub-problems: the \emph{Keen} and \emph{Act} tasks. The former aims to identify the set of items that a user is potentially interested in, while the latter aims to subsequently determine the set of activities to perform on the items of interest. An outline of our proposed \textsf{Keen2Act} model is given in Figure \ref{fig:framework}.

We first denote a particular user, item, and activity using the notation $u$, $v$ and $z$, respectively. We also let $U$, $V$ and $Z$ denote the set of all users, all items, and all activities, respectively. For a given user $u$, the Keen component determines whether an item $v$ should be included in the set $V^{+}_{u}$ of items selected by that user, as follows:
\begin{align}
V^{+}_{u} &= \left\{ v \in V: \mathcal{K}(u, v) \geq \delta_{\mathcal{K}}(v) \right\}
\end{align}
where $\mathcal{K}(u, v)$ is the Keen ranking score for user-item pair $(u, v)$, and $\delta_{\mathcal{K}}(v)$ is the Keen decision threshold for $v$. 
Subsequently, the Act component determines whether an activity $z$ should be part of the set $Z^{+}_{u,v}$ of activities performed by user $u$ on a selected item $v$:
\begin{align}
Z^{+}_{u,v} &= \left\{ z \in Z: \mathcal{A}(u, v, z) \geq  \delta_{\mathcal{A}}(z) \land v \in V^{+}_{u} \right\}    
\end{align}
where $\mathcal{A}(u, v, z)$ is the Act ranking score for the tuple $(u, v, z)$, and $\delta_{\mathcal{A}}(z)$ is the Act decision threshold for $z$. 

The main intuition behind the Keen followed by Act steps is that, prior to determining an activity $z$ on item $v$, user $u$ must be sufficiently keen in item $v$ in the first place. When there is a lack of keenness (i.e., $\mathcal{K}(u, v) < \delta_{\mathcal{K}}(v)$), the user should not perform any activity at all on item $v$. Conversely, only when the user shows a sufficient level of keenness (i.e., $\mathcal{K}(u, v) \geq \delta_{\mathcal{K}}(v)$), he/she can proceed with selecting which activities to be performed on item $v$. This two-stage decision process helps not only reduce the search space (by filtering out less relevant items) but also improve the quality of the final item-activity recommendation. 

The ranking scores $\mathcal{K}(u, v)$ and $\mathcal{A}(u, v, z)$ can each be realized using any machine learning model. In this work, we choose to use a multilinear model called \emph{Factorization Machine} (FM) \cite{Rendle2012}, which has shown good performance in a variety of recommendation tasks based on sparse data. It is also worth noting that the decision thresholds $\delta_{\mathcal{K}}(v)$ and $\delta_{\mathcal{A}}(z)$ are parameters that are learnable from data and are specific to each item $v$ and activity $z$, respectively. 

Finally, a \emph{Recommendation Aggregation} process takes place to combine $V^{+}_{u}$ and $Z^{+}_{u,v}$ in order to arrive at the final list $R^{+}_{u}$ of recommended item-activity pairs. More specifically, for a given user-item-activity triplet $(u, v, z)$, the Recommendation Aggregation process corresponds to a decision function $\mathcal{D}(u, v, z)$:
\begin{align}
\label{eqn:decision}
\mathcal{D}(u, v, z) &= \mathbb{I} \left[ v \in V^{+}_{u} \land z \in Z^{+}_{u,v} \right]
\end{align}
where $\mathbb{I}[.]$ is an indicator function and $\land$ is an AND logical operator.

We further elaborate the formulation of learning mechanisms behind the Keen and Act components in Sections \ref{sec:keen_model} and \ref{sec:act_model}, respectively. We then describe the Recommendation Aggregation process in greater detail in Section \ref{sec:aggregation}. Finally, we recap the overall \textsf{Keen2Act} learning procedure in Section \ref{sec:learning}.


\subsection{Keen Component}
\label{sec:keen_model}

\textbf{Keen ranking}.
We consider the problem of identifying the parameters $\theta_{\mathcal{K}}$ of the Keen model $\mathcal{K}(u, v)$ as a learning-to-rank task, and it is thus sensible to use a loss function optimized for ranking. Among the myriad of ranking loss functions, the \emph{Weighted Approximately Ranked Pairwise} (WARP) loss \cite{weston2011wsabie} in particular has been shown as a good criterion for recommendation tasks. An appealing trait is that WARP works for data that have only implicit feedback, making it well-suited to our problem. The key idea of WARP is that, for a given positive (observed) example, the remaining negative (unobserved) examples are randomly sampled until we find a pair of positive and negative examples for which the current model incorrectly ranks the negative example higher than the positive one. We can then perform a model update only based on these violating examples. 

Adopting WARP to the context of our Keen model, we can write the loss function with respect to the Keen model parameters $\theta_{\mathcal{K}}$ as:
\begin{align}
\mathcal{L}_\text{WARP}(\theta_{\mathcal{K}}) &= \sum_{u \in U} \sum_{v \in V^{+}_{u}} \Phi \left( rank_v(\mathcal{K}(u, v)) \right)
\end{align}
where $\Phi(.)$ transforms the rank of a positive item $v \in V^{+}_{u}$ into a weight. Here the $rank_v$ function can be defined as a margin-penalized rank of the following form:
\begin{align}
\label{eqn:warp_rank}
rank_v(\mathcal{K}(u, v)) &= \sum_{v' \notin V^{+}_{u}} \mathbb{I} \left( \mathcal{K}(u, v) \geq 1 + \mathcal{K}(u, v') \right)    
\end{align}

Choosing a transformation function such as $\Phi(n) = \sum_{i=1}^{n} \frac{1}{i}$ would allow the model to optimize for a better Precision@$k$. However, directly minimizing such a loss function via gradient-based algorithms would be computationally intractable, as equation (\ref{eqn:warp_rank}) sums over all items, resulting in a slow per-gradient update. We can circumvent this issue by replacing $rank_v$ with the following sampled approximation \cite{weston2011wsabie}: we sample $N$ items $v'$ until we find a violation, i.e, $\mathcal{K}(u, v) < 1 + \mathcal{K}(u, v')$, and subsequently estimate the rank, i.e., $rank_v(\mathcal{K}(u, v))$, by $\frac{|V \setminus V^{+}_{u}| - 1}{N}$. In addition to the WARP loss, we need an appropriate regularization term to control our model complexity. In this work, we employ L2 regularization, which is differentiable and suitable for gradient-based methods as well. This leads to an overall, regularized ranking loss function $\mathcal{L}_\text{rank}$:
\begin{align}
\label{eqn:keen_rank_loss}
\mathcal{L}_\text{rank}(\theta_{\mathcal{K}}) &= \mathcal{L}_\text{WARP}(\theta_{\mathcal{K}}) + \frac{\lambda_{\mathcal{K}}}{2}  || \theta_{\mathcal{K}} ||^{2}
\end{align}
where $\lambda_{\mathcal{K}} > 0$ is a user-specified $l_2$-regularization parameter. The term $\| \theta_{\mathcal{K}} \|^{2}$  is used  to  mitigate data overfitting by penalizing large parameter values, thus reducing the model complexity.

\textbf{Keen thresholding}.
Once the Keen ranking step is done, we need to determine the appropriate decision thresholds $\delta_{\mathcal{K}}$ in order to decide whether to include item $v$ into $V^{+}_{u}$. Ideally, we wish to identify a threshold such that the set of selected items matches the set of ground-truth, observed items as close as possible. However, it is difficult to learn the thresholds using such a set matching objective. We therefore relax this via the \emph{cross-entropy} loss function, and for the Keen model this would be:
\begin{align}
\label{eqn:keen_thres_loss}
\mathcal{L}_\text{thres}(\delta_{\mathcal{K}}) = \sum_{u \in U} \sum_{v \in V} CE \left( \mathcal{K}(u, v) - \delta_{\mathcal{K}}(v), I(v \in V^{+}_{u}) \right)
\end{align}
where $CE(x, y) = -\left[ y \ln(\sigma(x)) + (1 - y) \ln(1 - \sigma(x)) \right]$ is the cross entropy function, $\sigma(x) = \frac{1}{1 + \exp(-x)}$ is the sigmoid function, and $I(v \in V^{+}_{u})$ is an indicator function reflecting the ground truth for whether item $v$ belongs to the set of items selected by user $u$.

\subsection{Act Component}
\label{sec:act_model}

\textbf{Act ranking}. The ranking loss formulation for the Act model is similar to that of the Keen model except that the former deals with ranking of activities at the level of user-item pair. In particular, the WARP loss function associated with the Act model is given by:
\begin{align}
\mathcal{L}_\text{WARP}(\theta_{\mathcal{A}}) &= \sum_{u \in U} \sum_{v \in V^{+}_{u}} \sum_{z \in Z^{+}_{u,v}} \Phi \left( rank_z(\mathcal{A}(u, v, z)) \right)
\end{align}
where the rank function is likewise estimated by a sampled  approximation $rank_z(\mathcal{A}(u, v, z))$ $\approx \frac{|Z \setminus Z^{+}_{u,v}| - 1}{N}$. Accordingly, adding the L2 penalty to control the model complexity, the overall regularized WARP loss for the Act model $\mathcal{A}$ is given by:
\begin{align}
\label{eqn:act_rank_loss}
\mathcal{L}_\text{rank}(\theta_{\mathcal{A}}) &= \mathcal{L}_\text{WARP}(\theta_{\mathcal{A}}) + \frac{ \lambda_{\mathcal{A}}}{2}  || \theta_{\mathcal{A}} ||^{2}
\end{align}
where $\lambda_{\mathcal{A}} > 0$ is the L2-regularization parameter.



\textbf{Act thresholding}.
Similar to the thresholding in the Keen model, we can estimate the decision threshold of the Act model using the following cross-entropy loss:

\begin{align}
\label{eqn:act_thres_loss}
\mathcal{L}_\text{thres}(\delta_{\mathcal{A}}) = \sum_{u \in U} \sum_{v \in V^{+}_{u}} \sum_{z \in Z} CE \left( \mathcal{A}(u, v, z) - \delta_{\mathcal{A}}(z), I(z \in Z^{+}_{u,v}) \right)
\end{align}

where $I(z \in Z^{+}_{u,v})$ is an indicator function reflecting the ground truth for whether activity $z$ belongs to the set of activities carried out by user $u$ on a selected item $v$ (i.e., $v \in V^{+}_{u}$).

\subsection{Recommendation Aggregation}
\label{sec:aggregation}

The final step in the \textsf{Keen2Act} model is to generate a sorted list $R^{+}_{u}$ of recommended item-activity pairs for a given user $u$, by aggregating the selected item set $V^{+}_{u}$ and chosen activity set $Z^{+}_{u, v}$ as computed by the Keen and Act models respectively. To achieve this, we generate $R^{+}_{u}$ by enumerating the selected items $v \in V^{+}_{u}$ starting from the one with the highest $\mathcal{K}(u, v)$, and then for each selected item, we enumerate the selected activities $z \in Z^{+}_{u, v}$ starting from the highest $\mathcal{A}(u, v, z)$. This results in a recommendation list whereby the item ranking takes precedence over the activity ranking.

\subsection{Learning Procedure}
\label{sec:learning}

\begin{algorithm}[!t]
\footnotesize
\caption{\textsf{Keen2Act} Learning Procedure}
\label{alg:learning}
\begin{algorithmic}[1]
\Require
\Statex Keen interactions $\mathcal{I}_\mathcal{K}$ and Act interactions $\mathcal{I}_\mathcal{A}$
\Statex Maximum number of epochs $T$ and maximum negative samples $N$
\Ensure
\Statex Model parameters $\theta_\mathcal{K}$ and $\theta_\mathcal{A}$
\Statex Decision thresholds $\delta_\mathcal{K}$ and $\delta_\mathcal{A}$
\Statex \hrulefill
\Repeat \Comment{Phase 1: Rank learning}
    \For{each $(u, v) \in \mathcal{I}_{\mathcal{K}}$} \Comment{Keen ranking loop}
        \Repeat
            \State{Randomly sample a negative item $v'$ by from $V \setminus V^{+}_{u}$}
            \If{$\mathcal{K}(u, v) < 1 + \mathcal{K}(u, v')$} \Comment{Keen rank violation found}
                \State{Perform Adam update on $\theta_{\mathcal{K}}$ to minimise (\ref{eqn:keen_rank_loss})}
                \State{\textbf{break}}
            \EndIf
        \Until{maximum sampling $N$}
    \EndFor
    \For{each $(u, v, z) \in \mathcal{I}_{\mathcal{A}}$} \Comment{Act ranking loop}
        \Repeat
            \State{Randomly sample a negative activity $z'$ by from $Z \setminus Z^{+}_{u, v}$}
            \If{$\mathcal{A}(u, v, z) < 1 + \mathcal{A}(u, v, z')$} \Comment{Act rank violation found}
                \State{Perform Adam update on $\theta_{\mathcal{A}}$ to minimise (\ref{eqn:act_rank_loss})}
                \State{\textbf{break}}
            \EndIf
        \Until{maximum sampling $N$}
    \EndFor
\Until{maximum epochs $T$}
\Repeat \Comment{Phase 2: Threshold learning}
    \For{each user $u \in U$}
        \For{all items $v \in V$} \Comment{Keen thresholding loop}
            \State{Perform Adam update on $\delta_{\mathcal{K}}$ to minimise (\ref{eqn:keen_thres_loss})}
        \EndFor
        \For{all positive items $v \in V^{+}_{u}$} \Comment{Act thresholding loop}
            \For{all activities $z \in Z$}
                \State{Perform Adam update on $\delta_{\mathcal{A}}$ to minimise (\ref{eqn:act_thres_loss})}
            \EndFor
        \EndFor
    \EndFor
\Until{maximum epochs $T$}
\end{algorithmic}
\end{algorithm}

To minimize the ranking losses $\mathcal{L}_\text{rank}(\theta_{\mathcal{K}})$ and $\mathcal{L}_\text{rank}(\theta_{\mathcal{A}})$ as well as threshold losses $\mathcal{L}_\text{thres}(\delta_{\mathcal{K}})$ and $\mathcal{L}_\text{thres}(\delta_{\mathcal{A}})$, we devise an incremental learning procedure based on a variant of stochastic gradient descent called \emph{Adaptive Moment Estimation} (Adam) \cite{Kingma2015}. Algorithm \ref{alg:learning} summarizes the procedure, which takes in the Keen interactions $\mathcal{I}_{\mathcal{K}} = \{ (u, v): u \in U \land v \in V^{+}_{u} \}$ and Act interactions $\mathcal{I}_{\mathcal{A}} = \{ (u, v, z): u \in U \land v \in V^{+}_{u} \land z \in Z^{+}_{u,v}\}$ as inputs, and outputs the model parameters $\theta_\mathcal{K}$ and $\theta_\mathcal{A}$ as well as  thresholds $\delta_\mathcal{K}$ and $\delta_\mathcal{A}$. 

The overall algorithm consists of two phases: \emph{rank learning} and \emph{threshold learning}. In the first phase, we carry out Adam to update the model parameters $\theta_{\mathcal{K}}$ and $\theta_{\mathcal{A}}$ by finding a pair of positive and negative examples that violate the desired (Keen or Act) ranking. In the second phase, we also perform Adam update on  $\delta_{\mathcal{K}}$ and $\delta_{\mathcal{A}}$ by enumerating at the item and activity levels respectively.



\section{Experiment}
\label{sec:experiment}


\textbf{Datasets.} We experiment on two public datasets: GitHub \cite{Gousi13} and Stack Overflow\footnote{https://archive.org/details/stackexchange}. For both datasets, we retrieve active users (i.e., users who have performed at least 10 activities) and their activities performed between October 2013 to March 2015. For GitHub dataset, we obtain the \textit{fork} and \textit{watch} activities performed by 33,453 users on over 400k repositories. For Stack Overflow dataset, we retrieve the \textit{answer} and \textit{favorite} activities performed by 23,612 users on over 1 million questions. Table \ref{tbl:dataset} summarizes the two datasets.

\textbf{Features.} From the datasets, we construct two sets of features:
\begin{itemize}
	\item \textit{User features}: We adapt the \textit{co-participation similarity scores} introduced in \cite{lee2017github} to measure the level of co-participation between a given user and other users in the platform. For example, we compute the number of times two users \textit{co-fork} and \textit{co-watch} a GitHub repository. As such, each user is represented with a count vector with $d_U$ dimensions, where $d_U$ is equal to the number of users in the platform. 
	\item \textit{Item features}: For each item (i.e., Stack Overflow question and GitHub repository), we compute its TF-IDF vector based on its description tags. Hence, each item is represented with a TD-IDF vector with $d_T$ dimensions, where $d_T$ is equal to the total number of tags used to describe the items. 
\end{itemize}

\textbf{Baselines.}  As a few related works perform activity recommendations, we adapt and apply some of the commonly used item recommendation methods to our scenario. Specifically, we compare our model to two variants of Factorization Machine (FM):
\begin{itemize}
    \item \textbf{FM\_BPR} \cite{Rendle2012}. This method uses \emph{Bayesian Personalized Ranking} (BPR) loss \cite{Rendle2009} to maximize the rank difference between a positive example and a randomly chosen negative example. 
    \item \textbf{FM\_WARP} \cite{weston2011wsabie}. This method uses WARP loss to maximize the rank of positive examples by repeatedly sampling negative examples until a rank violation is found.
\end{itemize}
The user by item-activity interaction matrix, user, and item features are used as input for both baseline methods. All the parameters of baseline methods are empirically set to the optimal values. Besides the baselines, we also test several variants of our model:
\begin{itemize}
    \item \textbf{Keen Model}. Using only the Keen model, we retrieve a set of items that the user is interested in and recommend the user to perform all activities on the retrieved items.
    \item \textbf{Act Model}. Using only the Act model, we consider all possible activities for all possible user-item pair and recommend activities which meet the Act threshold.
    \item \textbf{\textsf{Keen2Act}}. Our full model with both Keen and Act modules.
\end{itemize} 


\begin{table}[!t]
\caption{Dataset summary}
\label{tbl:dataset}
\begin{tabular}{|l|c|l|c|}
\hline
\multicolumn{2}{|c|}{\textbf{GitHub}}  & \multicolumn{2}{|c|}{\textbf{Stack Overflow}} \\
\hline\hline
\#Users & 33,453 & \#Users & 23,612 \\
\#Repositories  & 461,931 & \#Questions  & 1,020,809 \\
\#Fork Activities & 445,084 & \#Answer Activities & 860,302 \\
\#Watch Activities& 1,730,181 & \#Favorite Activities & 544,617\\ \hline
\end{tabular}
\end{table}

\textbf{Training and testing splits}.
In all our experiments, we randomly select 80\% activities of each user to form the training set and use the remaining activities as testing set. As such, all users are observed in the training set. However, some items might be new in the test set (corresponding to a cold start problem). Note that the user and item features are computed based on the observations in the training set. We repeat this process 5 times, resulting in 5 training-testing splits based on which we evaluate our model.

\textbf{Evaluation metric}. 
We use the Mean Average Precision at top $k$ (MAP@$k$) 
as a primary metric in our experiments, which is popularly used to evaluate recommendation models \cite{tsoumakas2009mining}. We vary $k$ from $5$ to $\infty$ in order to examine the sensitivity of our model. Note that setting $k = \infty$ is equivalent to computing MAP.

\begin{table}[!t]
\caption{Experiment results of various methods}
\label{tbl:results}
\begin{tabular}{llccccc}
\hline
{\textbf{Dataset}} & {\textbf{Model}}  & {\textbf{MAP@5}} & {\textbf{MAP@10}} & {\textbf{MAP@20}} & {\textbf{MAP@50}} & {\textbf{MAP}} \\\hline\hline
GitHub  & FM\_BPR & 0.120 & 0.128 & 0.127 & 0.113 & 0.036 \\
        & FM\_WARP & 0.300 & 0.299 & 0.276 & 0.233 & 0.077 \\
        & Keen Model &0.298&0.297&0.269&0.224&0.074 \\  
        & Act Model &0.250&0.243&0.232&0.219&0.058 \\  
        & \textsf{Keen2Act} &\textbf{0.348}&\textbf{0.347}&\textbf{0.325}&\textbf{0.284}&\textbf{0.099} \\\hline 
Stack   & FM\_BPR & 0.103 & 0.109 & 0.108 & 0.100	                        & 0.033 \\
Overflow                & FM\_WARP & 0.180 & 0.183 & 0.177 & 0.160	& 0.050 \\
                & Keen Model &  0.238 & 0.234  &  0.230 & 0.209 & 0.054\\  
                & Act Model & 0.191 & 0.191 & 0.188 & 0.186 & 0.053 \\  
                & \textsf{Keen2Act} & \textbf{0.259} & \textbf{0.249}& \textbf{0.227} & \textbf{0.210}  &\textbf{0.064} \\\hline   
\end{tabular}
\end{table}

\subsection{Results and Discussion}

Table \ref{tbl:results} shows the MAP@$k$ results averaged over the 5 runs of our activity recommendation experiments. We observe that \textsf{Keen2Act} consistently outperforms all the other methods. Specifically, \textsf{Keen2Act} outperforms FM\_BPR and FM\_WARP by 175\% and 29\% respectively in Stack Overflow, and 94\% and 28\% respectively in GitHub. As we increase $k$, we also notice deterioration in MAP@$k$ results. This can be attributed to the Precision@$k$ metric favoring a model that outputs a ranked list with the relevant (i.e., observed) item-activity pairs leading the list. That is, as $k$ increases, it is  likely that more and more irrelevant item-activity pairs would appear in between the relevant item-activity pairs, pushing the Precision@$k$ lower. 

Additionally, we can see that the improvement of \textsf{Keen2Act} over the baselines is greater in Stack Overflow than in GitHub. A possible reason is due to the sparsity of user activities. More specifically, GitHub users are observed to perform more activities concentrated on a small set of items (i.e., denser user by item-activity interaction matrix), whereas Stack Overflow users tend to perform fewer activities spread across a large set of items. The denser interaction matrix for GitHub allows the baseline methods to have sufficient observations to achieve competitive results.

Comparing the different variants of our proposed model, we can see that \textsf{Keen2Act} outperforms the Keen model, suggesting that it is inadequate to learn only the item-level interests of the users when recommending activities. \textsf{Keen2Act} also outperforms the Act model, which demonstrates the importance of the Keen step in learning the item-level interests before activity-level interests. It is also interesting to see that the Keen model performs fairly well in comparison to the other methods. We can attribute this to reduced sparsity in the problem space it is operating at, i.e., the Keen model only makes item-level recommendation while the other methods recommend at the activity level. Finally, it is worth noting that the MAP@$k$ scores of various models are generally low, showcasing the complexity of the activity recommendation problem. 

\section{Conclusion}

In this paper, we put forward a new \textsf{Keen2Act} modeling approach to recommending items and multiple type of activity to a user in a step-wise manner. The efficacy of the approach has been demonstrated in experiments using real-world GitHub and Stack Overflow datasets.
In the future, we would like to extend \textsf{Keen2Act} using deep representation learning and conduct more comprehensive experiments to benchmark it against more state-of-the-art methods. We also plan to apply \textsf{Keen2Act} to other social platforms as well as consider a larger number of activity types.

\bibliographystyle{ACM-Reference-Format}
\bibliography{ref}

\end{document}